\documentclass[aip,graphicx,cha,amsmath,amssymb,preprint]{revtex4}
\usepackage{graphicx}% Include figure files
\usepackage{natbib}
\begin{document}
\title{Laminar-turbulent separatrix in a boundary layer flow}
\author{Damien BIAU}

\affiliation{Institut Pprime,\\
CNRS-Universit\'e de Poitiers-ENSMA,\\
T\'el\'eport 2-Bd. Marie et Pierre Curie B.P. 30179\\
86962 Futuroscope Chasseneuil Cedex, France}

\email{damien.biau@univ-poitiers.fr}
\date{29 March 2012}

\begin{abstract}
The transitional boundary layer flow over a flat plate is investigated. The boundary layer flow is known to develop unstable Tollmien-Schlichting waves above a critical value of the Reynolds number. However, it is also known that this transition can be observed for sub-critical Reynolds numbers. In that case the basin of attraction of the laminar state coexists with the sustained turbulence. In this article, the trajectory on the separatrix between these two states is simulated. The state on the separatrix is independent from the initial condition and is dynamically connected to both the laminar flow and the turbulence. Such an edge state provides information regarding the basic features of the transitional flow. The solution takes the form of a low speed streak, flanked by two quasi-streamwise sinuous vortices. The shape of the streaks is close to that simulated with the linear optimal perturbation method. This solution is compared to existing results concerning streak breakdown. The simulations are realized in a temporal framework for a local boundary layer, with periodic boundary conditions in the streamwise direction. A dedicated model, based on a scale separation, is presented.  The mean flow is a solution of the Prandtl boundary layer equations while the superposed small-scale fluctuations are a solution of the periodic Navier-Stokes equations. The model is validated with turbulent flow simulations and satisfactorily reproduces the physical characteristics of a boundary layer flow, especially in the outer region, where external fluid is entrained toward the boundary layer.
\end{abstract}

%\begin{keywords}
%\end{keywords}
%\setcounter{page}{1}
%\pacs{}% insert suggested PACS numbers in braces on next line
\maketitle %\maketitle must follow title, authors, abstract and \pacs

%--------------------------------------------------------------------------------------------------
\section{Introduction}
%--------------------------------------------------------------------------------------------------

%% generalities about turb. boundary layers
The boundary layer transition to turbulence is a long standing subject of research since it presents a good compromise between academic and industrial configurations. The delay of transition may be of applied interest, for example, to overcome turbulent aerodynamic drag. On the other hand, promoting transition permits an increase in mixing, heat transfer and other exchange processes. Since the first results by Lord Rayleigh, more than a century ago, this theoretical subject remains fascinating in its richness, as attested to by the regular number of related publications. It can be regarded as a benchmark problem for stability analysis. This simple shear flow also raises issues associated with, for instance, receptivity, chaos in open flows, modal versus algebraic growth, the emergence and the sustaining of coherent structures. 

%% various critical Re numbers: (neutral stability : $\alpha=0.303$ $\omega=0.12$ $Re_{\delta^*}= 519.4$)
The control parameter governing boundary layer transition is the Reynolds number, denoted Re, which can be defined in different ways. Hereafter the characteristic quantities are the boundary-layer displacement thickness $\delta^*$ and the free-stream velocity $U_\infty$, then $Re_{\delta^*}=U_\infty \delta^*/\nu$. The minimal Reynolds number below which disturbances decrease monotonically is found to be $Re_{\delta^*}= 17$\cite{footnote1}. The laminar state is globally stable, any finite-amplitude perturbations are monotonically decreasing. For $Re_{\delta^*}>519.4$, the laminar state is no longer an attractor and Tollmien-Schlichting (hereafter denoted to as T-S) travelling waves are amplified. In the range $17<Re_{\delta^*}<519.4$, the flow is conditionally stable, its stability depends on the disturbance shape and  energy. Infinitesimal perturbations can be transiently amplified and some coherent states, or traveling wave solutions, are expected to cohabit. Spalart \cite{Spalart} reported the possibility of sustained turbulence up to the minimum Reynolds number $Re_{\delta^*}= 400$ 
\cite{footnote2}. This is characteristic of the the sub-critical nature of the instability, indeed the nonlinear Tollmien-Schlichting waves, \emph{i.e.} the saturated T-S waves with a finite amplitude, can be prolonged below the critical value. For these two-dimensional equilibrium states Koch \cite{Koch} found the nonlinear critical Reynolds number: $Re_{\delta^*}=500$.

%% bypass transition, linear scenario
In addition to these non-linear aspects for the sub-critical transition, an important step was made, three decades ago, with the discovery of a possible transient  growth for the perturbation energy in a linear framework. While the classical analysis focuses on the  the asymptotic behavior determined by the least stable mode, a general, small disturbance is in fact a weighted combination of linear eigenfunctions. Because of the non-normality of the linearized stability operator, there is the potential for very large transient amplification of the disturbance energy, even in nominally stable flow conditions. For unstable flows the transient growth is able to overcome the classical T-S instability, in this case the transition is called bypass. This scenario is encountered for boundary layers subjected to moderate perturbations, for example a free stream turbulence with levels comprised between 1 and 6\% of the external velocity. Unlike the single mode case, the growth property depends on the disturbance environment, like surface roughness or free-stream turbulence. To overcome this difficulty the initial receptivity process can be studied using the optimal perturbation method : the inflow or initial condition is obtained through the maximisation of the transient energy growth. For the boundary layer, the temporal analysis was first performed by Butler \& Farrell \cite{ButlerFarrell} and extended later to the spatial non-parallel boundary layer by Andersson \emph{et al.} \cite{Andersson1} and Luchini \cite{L2000}. These works have shown the same features, the optimal perturbations take the form of damped streamwise vortices which generate streamwise streaks through a lift-up effect. The ratio between the input energy (vortices) and the output (streaks) is proportional to the Reynolds number: it is for this reason that the process is relevant to the bypass transition. Another bypass scenario is triggered by a pair of oblique waves, see Schmid \& Henningson \cite{Schmid_JFM_1992} and Berlin \emph{et al.} \cite{Berlin}. In this case, the streamwise vortices are generated by quadratic interaction between the oblique wave and, in turn, form the streamwise streaks. 
%Note that the bypass transition can be encountered in super-critical regime since the slow downstream exponential amplification of the T-S instability can be overwhelmed by the algebraic streaks instability.

%% non linear evolution
The linear mechanisms described so far are not sufficient to trigger transition in the sub-critical regime. In all cases, if the perturbations do not exceed some threshold, they eventually decrease and the flow state stays in the laminar basin of attraction. Thus the linear analysis has to be pursued by a weakly nonlinear or secondary stability analysis. The secondary instabilities can be differentiated following their spanwise symmetry, distinguishing the varicose (symmetric) from the sinuous (antisymmetric) structures. Again the streaks and T-S waves differ in their development. The nonlinear development of T-S waves instigates three dimensionality, giving rise to longitudinal streaks and varicose $\Lambda$-shaped vortices. These vortices can be aligned in the streamwise direction (K-type as reference to Klebanoff \emph{et al.} \cite{Klebanoff}) or staggered (H-type as reference to Herbert \cite{Herbert}). The late stage of oblique waves transition exhibits similar $\Lambda$ structures, see Berlin \emph{et al.} \cite{Berlin}. More details can be found in the book by Schmid \& Henningson \cite{YellowBook}. On the other hand, during the streak breakdown, computed by direct numerical simulations by Brandt \emph{et al.} \cite{Brandt2}, both sinuous and varicose modes are observed, with the former being more likely to occur. This feature is in agreement with various analyses: a modal inviscid stability by Andersson \emph{et al.} \cite{Andersson2}, non-modal growth by Hoepffner \emph{et al.} \cite{Hoepffner} and experimentally by Swearingen \& Blackwelder \cite{Swearingen}. While these studies show similar conclusions, it should be noted that the streaks considered are generated in different ways, by free-stream turbulence, as a result of optimal perturbations or through a Gortler instability. For further details, a recent overview of streak breakdown in bypass transition can be found in Schlatter \emph{et al.} \cite{Schlatter2}.

%  streaks and wall-cycle +mfu+ssp +TWS expe
A commonly observed feature at the onset of wall turbulence is the development of streamwise low-speed streaks. More precisely, the generic transitional state is composed of three main elements: the streamwise rolls sustain the streamwise streaks, by redistributing the mean momentum of the mean shear flow. Then the streaks become the support for wake-like instabilities, due to the spanwise or wall-normal shear, via modal or transient growth mechanisms, see Schoppa \& Hussain \cite{Schoppa}. Finally the nonlinear (quadratic) self-interaction of that streamwise fluctuations directly regenerates the streamwise rolls, closing the wall-cycle loop. 

For developed turbulence, such constituents are involved in a self-sustaining process and they represent the minimal self-sustaining flow unit capable of surviving viscous decay, as demonstrated for the plane Poiseuille flow by Jim\`enez \& Moin \cite{Jimenez}.
The minimal flow unit corresponds to the smallest computational box in which the turbulence may be sustained. For each wall layer, a single set of coherent structures was identified, composed of a pair of high and low-speed streaks with a pair of counter-rotating, quasi-streamwise vortices. The application to the plane Couette flow allowed Hamilton \emph{et al.} \cite{Hamilton} to define the  self-sustaining process as a quasi-cyclic dynamic, comprising quiescent periods interspersed by bursts of shorter duration. 

While the coherent structures were previously obtained by continuation of a sub-critical instability by Koch \cite{Koch}, or by continuation from an unstable brother flow (for example from Taylor-Couette to plane Couette flow, see Nagata \cite{Nagata}), the paradigm of the self-sustaining process provides a method by which to calculate the exact coherent structures by continuation of solutions that bifurcate from a streaky flow as explained in Waleffe \cite{Waleffe}. Other such solutions were later found for plane channel flow and pipe flows. As for the solution continued from the marginal instability, they survive down to Reynolds numbers below the critical value for turbulence onset. These results suggest that the underlying process is generic and fundamental to both developed turbulence and transition, see Wang \emph{et al.} \cite{Wang}. The contribution of Waleffe \cite{Waleffe2} should be mentioned here: he described the unity between turbulence in the minimal flow unit, the self-sustaining process of the developed turbulence and the role of the embedded traveling wave solutions. It is not the aim here to provide a complete bibliography; overviews are provided by Cvitanovic \& Gibson \cite{Cvitanovic}, Kawahara \cite{Kawahara2} and Panton \cite{Panton}. Beyond the various theoretical approaches, a promising and exciting result is the direct observation of these traveling waves in the pipe flow experiments by Hof \emph{et al.} \cite{Hof}. For spatially developing flows, Duriez \emph{et al.} \cite{Duriez} demonstrated, experimentally, the relevance of the self-sustaining process in a boundary layer. By varying the amplitude of a vortex generator,\emph{i.e.} small cylinders, the results show a threshold above which the streamwise vortices are regenerated from the destabilization of the streaks.

%%  the edge state and the numerical issue
The minimal flow unit was introduced by Jim\`enez \& Moin \cite{Jimenez} as a tool for studying near wall turbulent flow with a lower complexity. In the sub-critical case, another strategy for minimizing chaotic activity consists in maintaining the trajectory on the separatrix, namely the surface separating the laminar basin of attraction and the turbulent trajectories. A set of initial states that dynamically evolve to the laminar attractor forms its basin of attraction, which is separated from the rest of the state space by the separatrix. With a repeated bisection method it is possible to maintain the transitional state, for arbitrary long times, on the separatrix. This method was initially proposed by Toh \& Itano in 2003 \cite{Toh}, who computed a periodic-like trajectory of channel flow. The edge state maintained on the separatrix is called the edge of chaos by Skufca \emph{et al.} \cite{Skufca}. The method was applied latter by Schneider \emph{et al.} \cite{Schneider2} to a pipe flow and by Schneider \emph{et al.} \cite{Schneider1} to a plane Couette flow. It is expected that knowledge of these finite-amplitude solutions will contribute much to the unraveling of the structure of  transitional turbulent flows.

In this article we are interested in the transitional boundary layer for the sub-critical regime. The main objective is to compute a sustained solution, dynamically connected to the laminar and the turbulent states. The flow configuration is chosen such that the two attractors coexist and the trajectory is maintained on the separatrix with a bisection method. The results are compared to the various linear and nonlinear scenarios, mentioned above, describing the sub-critical laminar-turbulent transition. Whereas the path to turbulence depends on the shape and the amplitude of the initial condition, in contrast, the converged trajectory on the separatrix does not depend on its initial condition. Thus the corresponding edge state can provide some general features of transitional states.

A secondary objective concerns the numerical issues associated with the simulation of boundary layer flows, since the previous work for simulating the separatrix were realized for parallel flows. In order to extend the method to the nearly parallel boundary layer flow, it is necessary to keep the periodic condition in the streamwise direction. However, a temporal boundary layer is not satisfactory since the boundary layer thickness increases in time. In the next section a model is presented to simulate a local boundary layer, namely with a constant displacement thickness ($\delta^*$); this model owes much to the previous work by Spalart \cite{Spalart}.

%--------------------------------------------------------------------------------------------------
\section{Method}\label{sec:method}
%--------------------------------------------------------------------------------------------------
The overall goal of this section is to obtain a model, with periodic boundary condition in the streamwise direction, which can provide an approximate solution of the local boundary layer. The coordinates and velocities are made non-dimensional, respectively, using the boundary-layer displacement thickness $\delta^*=\int_0^\infty 1-U~dy$ and the free-stream velocity $U_\infty$. The associated Reynolds number is $Re_{\delta^*}=U_\infty \delta^*/\nu$.

To retain the advantages of the spectral Fourier decomposition in the streamwise direction, it is possible to add a fringe region downstream of the physical domain. The outflowing fluid is constrained, via a volume force (which is significantly non-zero only within the fringe region), to the desired inflow condition, see for example Brandt \emph{et al.} \cite{Brandt}. The problem of specifying the inflow boundary condition remains. Lund \emph{et al.} \cite{Lund} suggested that the velocity field at a downstream station be rescaled and re-introduced at the inlet. The rescaling is  based on the similarity laws of the boundary layer, the difficulty comes from the two different wall-normal length scales : the law of the wall in the inner part and the defect law in the outer part of the boundary layer. This two-layer nature of the turbulence introduces spurious periodicity into the time series during the recycling (see Siemens \emph{et al.} \cite{Simens}). As a consequence, a substantial initial part of the computational box has to be discarded. The numerical cost leads us to reject this method in our case. An alternative consists in simulating a temporal boundary layer in a moving frame of reference, see Spalart \& Yang \cite{SpalartYang}. However the undisturbed boundary layer is not a solution of the unsteady boundary layer equations. In addition, the boundary layer thickness increases in time so the base flow eventually becomes unstable, thus it is not possible to reconcile the existence of an edge state with the large time necessary to obtain converged results. Thus it appears necessary to introduce a specific method for a local turbulent boundary layer. Following the method introduced by Spalart \cite{Spalart}, the velocity field $ \mathbf{u}$ is decomposed as a local boundary layer profile $U$ plus the deviation: 
\begin{equation}
 \mathbf{u}(x,y,z,t)=\mathbf{U}(x,y,t)+ \mathbf{u'}(x,y,z,t). 
\label{equ0}
\end{equation}

The mean flow is solution of the turbulent boundary layer equations:
\begin{equation}
\begin{array}{l}
\displaystyle  \frac{\partial U}{\partial x}+\frac{\partial V}{\partial y}  =  0 \\[8pt]
\displaystyle  \frac{\partial U}{\partial t} + \left( U\frac{\partial U}{\partial x}+V\frac{\partial U}{\partial y}\right)/Re   =  
\frac{1}{Re}\frac{\partial^2 U}{\partial y^2} -\frac{\partial \overline{u'v'}}{\partial y}
\end{array}
\label{equ2}
\end{equation}

The overbar on the Reynolds stress indicates a streamwise and spanwise averaging. The momentum equation is Reynolds number dependent because the time scale is of order $\delta^*/U_\infty$, in agreement with the time scale of the streamwise fluctuations (see equations \ref{equ1}). One major consequence of the non parallel effect is the continuous entry of external irrotational flow while the boundary layer thickness increases downstream. To take into account this effect, a new coordinate is introduced : $\eta=y/\delta^*$ and the contravariant velocities (letters topped by a tilde) are used:

\begin{equation}
\begin{array}{l}
\displaystyle  \tilde{U} = U\\[8pt]
\displaystyle  \frac{\partial \tilde{U}}{\partial x}=\frac{\partial U}{\partial x}+\frac{d \eta}{dx}~y~\frac{\partial U}{\partial y}\\[8pt]
\displaystyle  \tilde{V} = V - \frac{d \eta}{d x}~ y~ U 
\end{array}
\label{equ3}
\end{equation}
The wall-normal derivatives, $\partial \tilde{U}/\partial y$ and $\partial \tilde{V}/\partial y$ are computed directly from $\tilde{U}$ and $\tilde{V}$. This transformation is classical for adapted meshes, see also Spalart \cite{Spalart}. With this transformation $\tilde{V}$ is zero in the wall layer and becomes negative farther from the wall, thus the non-turbulent, or irrotational, fluid is entrained into the boundary layer.

%$$V_\infty (X_2-X_1) = \int_\delta^\infty U ~ dy$$

The boundary layer equations are solved between two locations: $X_1$ and $X_2$, for an infinitesimal distance $X_2-X_1$. In this sense the spatial boundary layer is simulated locally, for a given $Re_{\delta^*}$. This distinguishes the present model from the temporal boundary layer flow. The former describes, locally, the temporal evolution of a spatial boundary layer (\emph{i.e.} fixed $\delta^*$) while the latter describes a  temporal boundary layer flow (\emph{i.e.} with $\delta^*$ evolving in time).

At the location $X_1$, the displacement thickness $\delta^*(X_1)$ is determined, through a shooting method, such that $\delta^*(X_2)=1$. Now that $\eta$ is defined, the streamwise velocity at $X_1$ is determined, assuming a self-similar behavior :
\begin{equation}
U(X_1,~ \eta)=U(X_2,~ y/\delta^*(X_1))
\label{equ_scaling}
\end{equation}

It is now possible to compute the streamwise derivatives, in equations \ref{equ2} and \ref{equ3}, by taking the differences :
%$$\frac{d \eta}{dx}=\frac{1-1/\delta^*(X_1)}{X_2-X_1}\quad \mathrm{and} \quad 
%\frac{\partial U}{\partial x}=\frac{U(X_1,~ y/\delta^*(X_1))-U(X_2,~ y)}{X_2-X_1}$$
\begin{equation}
\begin{array}{l}
\displaystyle  \frac{d \eta}{d x}=\frac{y-y/\delta^*(X_1)}{X_2-X_1}\\[8pt]
\displaystyle  \frac{\partial U}{\partial x}=\frac{U(X_1,~ y/\delta^*(X_1))-U(X_2,~ y)}{X_2-X_1}
\end{array}
\label{equ4}
\end{equation}

The results appear to be weakly sensitive to the spacing $X_2-X_1$ provided it is sufficiently small, at least lower than $0.5$. In the following, the simulations are realized with $X_2-X_1=10^{-2}$.

The deviation is assumed to evolve on a short scale around a weakly nonparallel mean flow evolving on a long scale. The equation for the deviation truncated at order $1/Re_{\delta^*}$ becomes: 

\begin{equation}
\begin{array}{l}\displaystyle 
\frac{\partial u'}{\partial x}+\frac{\partial v'}{\partial y}+\frac{\partial w'}{\partial z}  =  0 \\[8pt]
\displaystyle 
\frac{\partial u'}{\partial t} + \left(\tilde{U}+u'\right)\frac{\partial u'}{\partial x}
+\left(\tilde{V}/Re+v'\right)\frac{\partial u'}{\partial y} 
+\frac{\partial \tilde{U}}{\partial y} v'+\frac{\partial \tilde{U}/Re}{\partial x}~ u'+ w'\frac{\partial u'}{\partial z}
=-\frac{\partial p'}{\partial x} + \Delta u'/Re \\[8pt]
\displaystyle 
\frac{\partial v'}{\partial t} + \left(\tilde{U}+u'\right)\frac{\partial v'}{\partial x}
+\left(\tilde{V}/Re+v'\right)\frac{\partial v'}{\partial y} 
+ \frac{\partial \tilde{V}/Re}{\partial y} ~v'   +w'\frac{\partial v'}{\partial z} 
=  -\frac{\partial p'}{\partial y} + \Delta v'/Re \\[8pt]
\displaystyle 
\frac{\partial w'}{\partial t} + \left(\tilde{U}+u'\right)\frac{\partial w'}{\partial x}
+\left(\tilde{V}/Re+v'\right)\frac{\partial w'}{\partial y} +w'\frac{\partial w'}{\partial z}
=  -\frac{\partial p'}{\partial z} + \Delta w'/Re 
\end{array}
\label{equ1}
\end{equation}
\noindent with a compact form of the Laplacian operator $\Delta=\partial^2/\partial x^2+\partial^2/\partial y^2+\partial^2/\partial z^2$. 
These equations are associated with homogeneous boundary conditions at the wall and at the upper boundary. The results appear not to be influenced by the external homogeneous boundary condition since the simulated fluctuations are irrotational and behave asymptotically as 
$\exp(-k~y)$ with $k=\sqrt{\alpha^2+\beta^2}$.

The two strong assumptions are the periodicity applied to streamwise (\emph{i.e.} x-independent) streaks and vortices in equation \ref{equ1} and the rescaling of the mean streamwise velocity.  On the other hand, the model is self-governing, which means there is no need for an empirical adjustment with previously existing data. The model could be interesting in order to generate an inflow condition for a spatially developing turbulent boundary layer.

The equations are discretized using a Chebyshev collocation method in the wall-normal direction ($y$) and Fourier in the streamwise ($x$) and spanwise ($z$) direction. In the wall-normal direction, the Gauss-Lobatto-Chebyshev points ($\zeta$) are expanded from $[-1,~ 1]$ to 
$[0,~ y_\infty]$ throughout an algebraic mapping: $y = (y_\infty h)/2~ (1+\zeta)/(1+h−\zeta)$. Where $h=0.5$ is a stretching factor and $y_\infty=15$ is the location of the upper boundary. The differentiation operators are suitably adjusted by the analytical Jacobian of the mapping. Integral quantities are obtained using similarly-adjusted Gaussian quadrature rules. For the time-integration an order three semi-implicit finite difference scheme is used, associated with a prediction/projection method. In order to compute a pressure unpolluted by spurious modes, the pressure is approximated by polynomials ($P_{N-2}$) of two units lower-order than for the velocity ($P_N$), see Botella \emph{et al.} \cite{Botella}.

%In order to validate the model, two simulations of turbulent boundary layers have been realized. 
The simulation must resolve the entire range of scales of the turbulent motion : from the boundary layer thickness down to the dissipative scale. As an indication, the mean spanwise spacing of the streaks in the buffer layer ($y^+<30$) is about 100 wall units, the superscript + designates quantities expressed in the wall units based on the kinematic viscosity $\nu$ and the skin friction velocity $u_\tau$. The first constraint implies a large domain while the latter imposes a large number of grid points. The period in the streamwise and spanwise directions are $L_x=50$ and $L_z=15$. In the transverse directions the grid spacing is $\Delta z^+\approx 6$ and $\Delta x^+\approx 20$, while in the wall-normal direction, at least 14 points are within 9 wall units of the wall. These requirements match those by Spalart \cite{Spalart}, except that the domain is slightly shorter in the present simulations. The associated number of points in the x-, y-, and z-directions is $(64\times 75 \times 64)$ at $Re_{\delta^*}=500$ and $(128\times 95 \times 128)$ at $Re_{\delta^*}=1000$, while the timesteps  are $\Delta t=0.1$ and $\Delta t=0.05$, respectively.

\begin{figure}[!h]
  \centerline{\includegraphics[width=10cm]{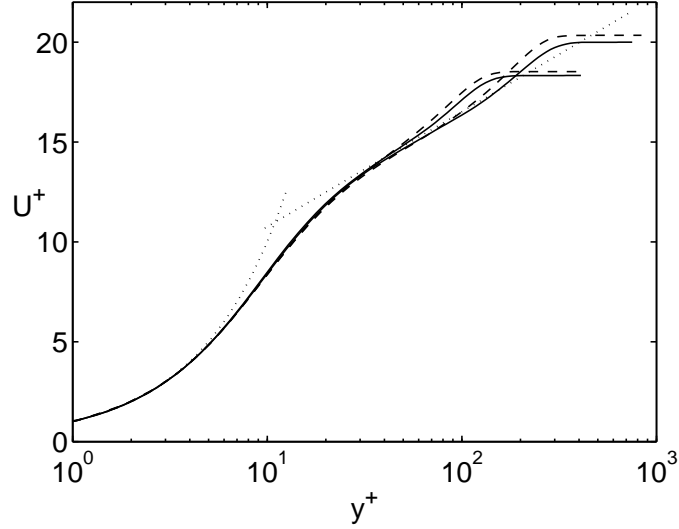}}
  \caption{Mean velocity profiles at $Re_{\delta^*}=500$ (bottom curves) and $Re_{\delta^*}=1000$ (upper curves).
Continuous lines :  present results, dashed lines Spalart \cite{Spalart}. 
The dotted line indicate the classical viscous sublayer $y^+$ and the log law $\kappa^{-1}\ln y^+ + C$, with $\kappa=0.4$ and $C=5$.}
 \label{turbU}
\end{figure}

For zero fluctuations, the laminar Blasius flow is reproduced with an absolute error lower than $0.15\%$. The model is also able to  correctly reproduce the linear growth of Tollmien-Schlichting waves. The ability of the model to reproduce the turbulent boundary layer now needs to be demonstrated by comparison with other results. The mean velocity profiles are displayed in figure \ref{turbU}.

The agreement is correct in the buffer layer ($y^+<30$), an area in which the DNS is very reliable. Just above, in the log layer, the agreement is also satisfactory but farther from the wall, in the wake region, the flow is misrepresented as expected from a nearly parallel model. The skin friction velocity is overestimated for both nearly parallel simulations.

\begin{figure}[!h]
  \centerline{\includegraphics[width=10cm]{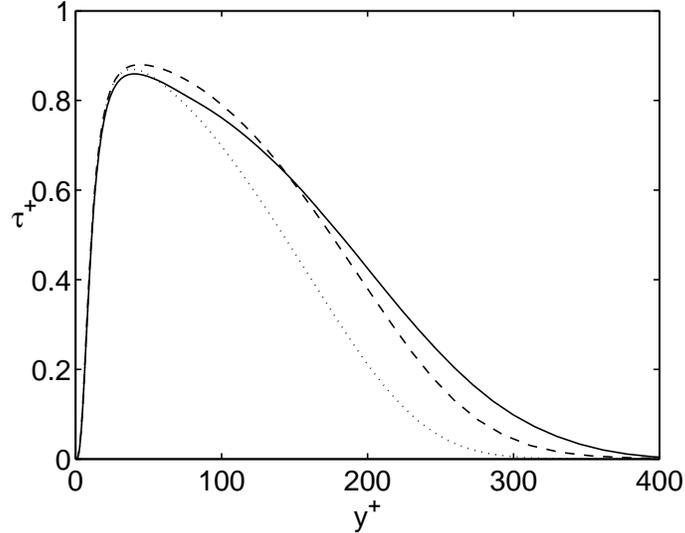}}
  \caption{Reynolds stress distribution, normalized by the friction velocity, at $Re_{\delta^*}=1000$. 
Continuous line : present results, dashed line Spalart \cite{Spalart}, dotted line Schlatter \& \"Orl\"u \cite{Schlatter}}.
 \label{turbuv}
\end{figure}

This effect is more pronounced for the Reynolds shear stress ($\tau=-\overline{uv}^+$), displayed in figure \ref{turbuv}, which is wider than expected. The slight differences between present results and those by Spalart \cite{Spalart} can be imputed to the stronger parallel-flow hypothesis since the streamwise streaks and rolls are rescaled in the results by Spalart \cite{Spalart}.

To conclude with quantitative comparisons, the Reynolds number based on the momentum boundary layer thickness is $Re_\theta=307$, for the case $Re_{\delta^*}=500$, which is in agreement with the value $Re_\theta=303$ computed by Spalart \cite{Spalart}. For the case $Re_{\delta^*}=1000$ the present result is $Re_\theta=683$, in relatively good agreement with the values 677 found by Spalart \cite{Spalart} and 670 found by Schlatter \emph{et al.} \cite{Schlatter}. The fluctuating wall-shear stress are $\tau_{w,rms}=0.372$ and $0.3922$ for $Re_{\delta^*}=500$ and $1000$, respectively. These values are close to the classical value of $0.4$, also in good agreement with the relation given by Alfredsson \emph{et al.} \cite{Alfredsson} and \"{O}rl\"{u} \& Schlatter \cite{Orlu}: $\tau_{w,rms}=0.298+0.018 \ln Re_\tau$ leading to: $\tau_{w,rms}=0.388$ and $0.401$.

However, we will see that the separatrix is close to the laminar state, in particular the Reynolds shear stress for the edge state is three time smaller than for the turbulent state. In addition, the turbulent inner layer being absent, the mean flow $U$ scales exactly with the outer scale $\delta^*$, in agreement with the model (see equation \ref{equ_scaling}). Bearing in mind that the main issue of the paper is the understanding of the edge state and not the prediction of turbulent state, these results are deemed sufficient validation of the code for the present purpose.

A trajectory on the separatrix is bounded by initial conditions which eventually decay or become turbulent. Practically, we consider a turbulent velocity field $\mathbf{u}$, decomposed as shown in equation \ref{equ0}, with weighted fluctuations:
\begin{equation}\label{EdgeStateEq}
 \mathbf{u}(x,y,z,t)=\mathbf{U}(x,y,t)+ \lambda \mathbf{u'}(x,y,z,t). 
\end{equation}
For values of the weight $\lambda$ close to unity, the flow stays turbulent. By reducing $\lambda$, the initial condition moves closer to the laminar profile and will not become turbulent again. Therefore an interval of $\lambda$ values can be found, bounded at one end by a trajectory that becomes turbulent and at the other by a trajectory that returns to the laminar state. In between two such trajectories lies one  which resides for a substantial time interval on the separatrix. Thus the numerical method is based on a bisection algorithm in order to find the amplitude of two initial conditions on either side of the edge state. The bisection algorithm is initialised with a turbulent flow but the choice of initial state is not critical since  arbitrary initial conditions appear to converge to the same solution.
This edge state is only a partial attractor, equivalent to a saddle point in a phase space. Since it is necessary to repeat this procedure (after every 200 time units in our case) to constrain the solution on the separatrix and then to obtain a converged solution. After convergence, the coefficient $\lambda$ is very close to unity, with an accuracy of five decimals places, so the edge state can be considered as a solution of the nonlinear Navier-Stokes equations. 

It should be noted that this method is different from finding a threshold for transition to turbulence for a given initial condition. In the latter case the trajectory is found close to the separatrix but it is not exactly on the separatrix. Although there are several paths from the laminar state to turbulence, depending on the initial condition, the state on the separatrix is independent on the initial condition.
In other words, the laminar basin of attraction seems to be simply connected.

\begin{figure}[!h]
\centerline{\includegraphics[width=10cm]{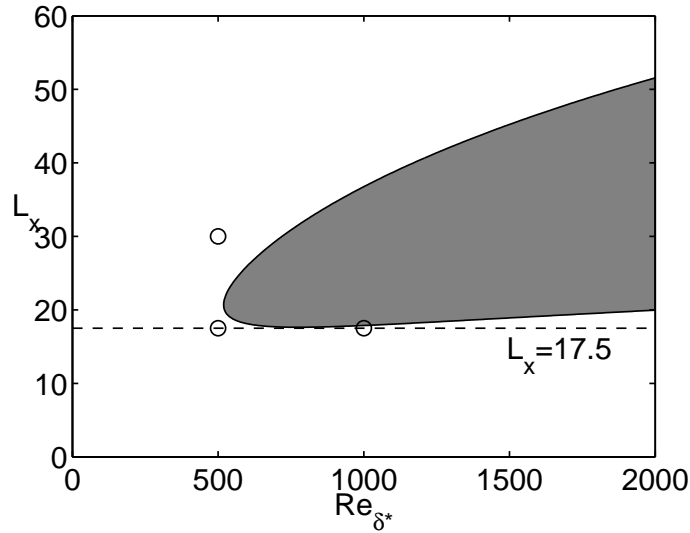}}
\caption{\label{neutral} Neutral stability curve for the Tollmien-Schlichting perturbation : Reynolds number ($Re_{\delta^*}$) versus  streamwise length ($L_x$), the unstable domain is the gray area. For streamwise length shorter than $L_x=17.5$ the Blasius boundary layer is unconditionally stable. The three circles represent the parameters used for the simulations.}
\end{figure}

Even if the basin of attraction of the laminar state disappears for $Re_{\delta^*}>519.4$, the neutral stability curve, in the figure \ref{neutral}, shows that for streamwise length shorter than $L_x=17.5$ the Blasius boundary layer is unconditionally stable and the laminar state is always surrounded by a basin of attraction, for all Reynolds numbers.

Thus it becomes possible to investigate the Reynolds number effect, even for super-critical values, by using a streamwise length equal to 17.5. Eventually three sets of parameters have been retained.
The main simulation was realized with $Re_{\delta^*}=500$ with $L_x=30$. Two complementary simulations have been realized with $L_x=17.5$ for two Reynolds numbers: $Re_{\delta^*}=500$ and $1000$.  In all simulations the numerical parameters are $N_x=64,~ N_y=75,~ N_z=64$ and $\Delta t=0.05$.

In order to check the improvement of the model presented in this section, a preliminary simulation has been performed with a parallel Blasius solution as a mean flow. The parallel assumption leads to numerical issues, the perturbations extend outside the boundary layer and eventually fill all the numerical domain, as for a half-channel flow.

%--------------------------------------------------------------------------------------------------
\section{Results}\label{sec:results}
%--------------------------------------------------------------------------------------------------

\begin{figure}[!h]
\centerline{\includegraphics[width=12cm]{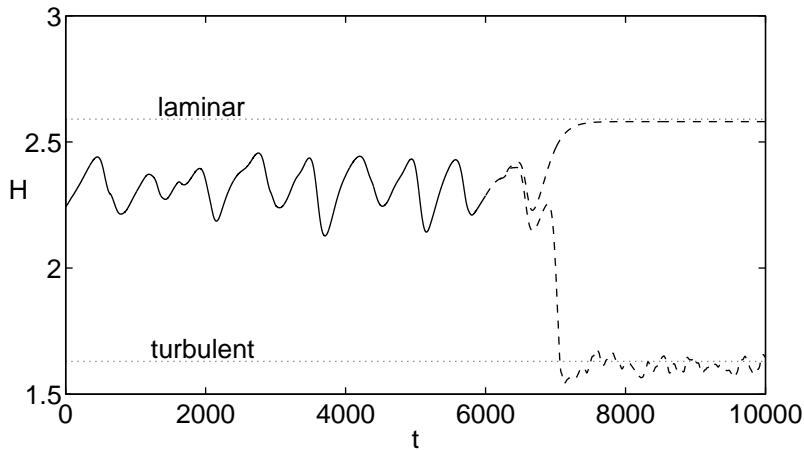}}
\caption{\label{EdgeState_H} Shape factor $H$ versus time for $Re_{\delta^*}=500$ and $L_x=30$. Dotted lines indicate the laminar and the (time-averaged) turbulent values. The continuous line corresponds to the separatrix, the two dashed lines represents the trajectories on either side of the edge state.}
%The circles indicate the instants where the bisection method is applied to maintain the trajectory on the separatrix}
\end{figure}

%  the shape factor --------------
Firstly in order to illustrate the dynamic on the separatrix, the time evolution of the shape factor is depicted on the figure \ref{EdgeState_H}. The shape factor $ H=\delta^*/\theta$, with $\theta$ the momentum thickness, is often used to determine the nature of the boundary layer flow. The time evolution of the shape factor in figure \ref{EdgeState_H} indicates that the separatrix is much closer to the laminar state ($H=2.59$) as compared to the turbulent state ($H=1.63$).

The bisection method is stopped after 6000 time units. After $t=6000$ the trajectory is ejected away in two possible directions: toward the laminar attractor or toward the turbulence, plotted with dashed lines in the figure \ref{EdgeState_H}. The laminar and  turbulent trajectories remain close for 500 time units, confirming that the numerical solution is very close to the separatrix. When the trajectory escapes from the separatrix, along the laminar path, the streamwise fluctuations disappear first and the flow tends to be formed by quasi-straight rolls and streaks, which then decay on a (slow) viscous time scale towards the laminar fixed point.  It should be noted that the discrepancy between the final value ($H=2.58$) and the expected Blasius value ($H=2.591$) can be explained by the approximated Blasius solution computed with equations \ref{equ2}. On the other hand, during the transition to turbulence the streamwise streaks and vortices move closer and closer to the wall, prefiguring the developed turbulence where the streamwise vortices populate the inner region; however the turbulent streaks are not an extension of the transitional streaks. Thus the two-layer nature of the wall-turbulence emerges: close to the wall, energy production exceeds dissipation, a part of this energy is exported far from the wall where dissipation dominates. In between a region  develops, comparable to the inertial range of scales in the energy cascade of isotropic turbulence, where the production balances the dissipation. But this cannot be considered as a logarithmic region for this low Reynolds number.

\begin{figure}[!h]
\centerline{\includegraphics[width=16cm]{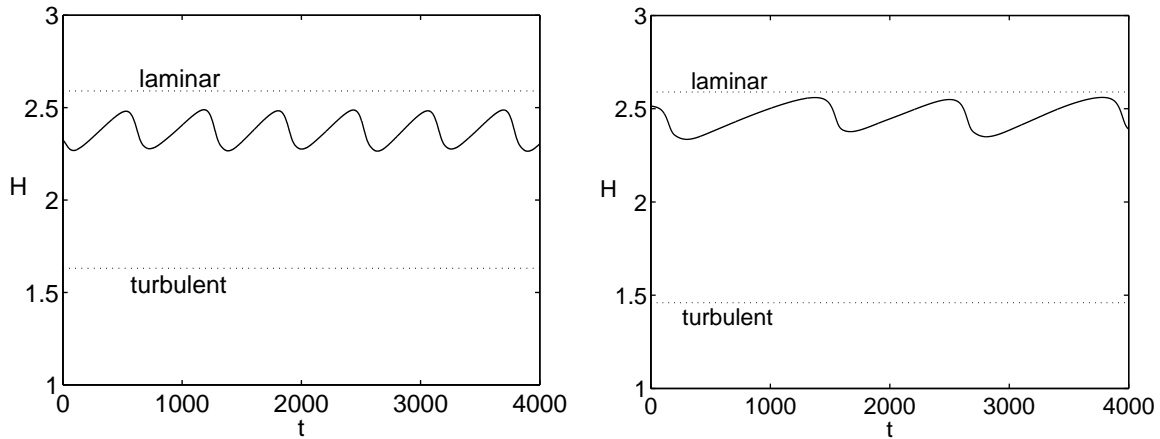}}
%H_Re0500_Lx17.eps}\includegraphics[width=8cm]{FIG/H_Re1000_Lx17.eps}}
\caption{\label{EdgeState_H2} Shape factor $H$ versus time for $L_x=17.5$ with $Re_{\delta^*}=500$ (left) and $Re_{\delta^*}=1000$ (right).}
\end{figure}

The case with higher Reynolds number $Re_{\delta^*}=1000$ and shorter streamwise length $L_x=17.5$ is plotted in figure \ref{EdgeState_H2}.
The dynamic constrained in a short domain seems to be a periodic limit cycle. The period is found to be almost constant and close to $1200$.
The simulation with a lower Reynolds number ($Re_{\delta^*}=500$) presents the same oscillations with a period of about 630. However, as stated by Toh \& Itano \cite{Toh}, we cannot conclude whether the periodic-like solution is a heteroclinic cycle or an exact periodic solution. On the other hand, for pipe flow, Duguet \emph{et al.} \cite{Duguet} indicate that trajectories on the laminar-turbulent boundary could be organized around only a few traveling wave solutions and their heteroclinic connections. See also Cvitanovic  \& Gibson \cite{Cvitanovic} for the relations between the periodic solutions and heteroclinic cycle in plane Couette flows. In the present case, the cyclical behavior seems to be a numerical artifact induced by a short streamwise length. In the following we focus on the results obtained for the larger box $L_x=30$ and a sub-critical Reynolds number $Re_{\delta^*}=500$.

\begin{figure}[!h]
\centerline{\includegraphics[width=14cm]{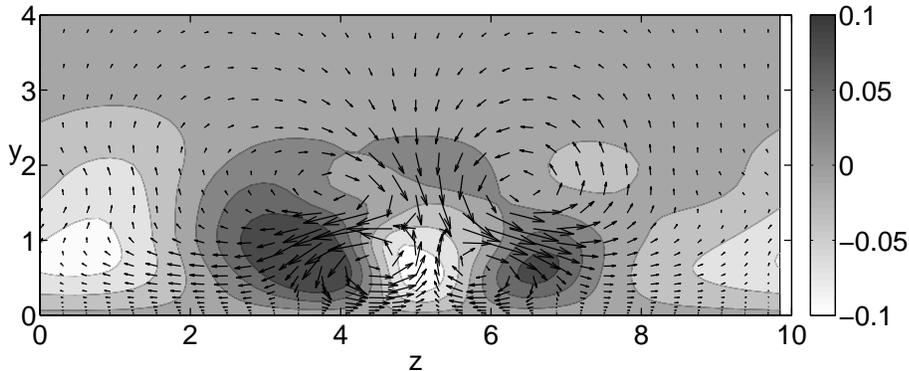}}
\caption{\label{EdgeState_Streaks_Rolls} Representation of the streamwise averaged fluctuations at $t=2800$ for $Re_{\delta^*}=500$ and $L_x=30$. The cross-stream velocity components (vortices) are  depicted with vectors while the streaks are identified by the grayed areas.}
\end{figure}

%  streaks and vortices ------------  Niter=5 t=2800
When numerical convergence is reached, the edge state does not depend on its initial condition. In that sense it is interesting to compare this state with the last state of transitional flow presented in various other investigations. Figure \ref{EdgeState_Streaks_Rolls} shows a cross section of the flow field at $t=2800$. Streamwise vortices are associated with low and high speed streaks throughout the lift-up process describing the linear interaction between the rolls and the mean shear flow. The non-linear interaction between the rolls and the streaks explains why the width of the high speed streaks is larger than that of the low speed streaks: the high speed streak, induced by a negative vertical velocity, is moved closer to the wall and spreads in the spanwise directions.

The most energetic structure is dominated by the low-speed streak associated with a pair of strong counter-rotating vortices near the wall, and above a pair of opposite counter-rotating vortices;  these are similar to those found in a circular pipe flow by Schneider \emph{et al.} \cite{Schneider2} or in square duct by Biau \& Bottaro \cite{Biau}. This structure, with double-layer vortices is a tenuous equilibrium between two antagonistic phenomena: the ejection and sweep events. The low speed streak, generated by the two near-wall vortices, is stretched in the spanwise direction and compressed in the wall-normal direction. It is the location of large, potentially inflectional, shear stress. This tenuous equilibrium can explain the periodic-like behavior mentioned before. One cycle consists of two typical intervals: a single-streak development and its sudden slip in the spanwise direction. The kinetic energy of the fluctuations is dissipated gradually in the former case, while it is recovered quickly in the latter case. This cycle is present in each case but it is less regular for the case with larger streamwise length.

%The time evolution of the streaks meandering is shown in the figure \ref{meandering}. 
%\begin{figure}[!h]
%\centerline{\includegraphics[width=10cm]{FIG/meandering.eps}}
%\caption{\label{meandering} Contours of the streamwise averaged streaks versus time, at a distance $y=1$ from the wall for 
%$Re_{\delta^*}=500$.}
%\end{figure}

%The shape factor decrease in the figure \ref{EdgeState_H} corresponds to a sweep event. The low speed streak (light gray) is pushed towards the wall, then splits into two low-speed streaks and a high speed streak develops between them, leading to a  slip overall movement in the spanwise direction.

\begin{figure}[!h]
\begin{center}
\includegraphics[width=10cm]{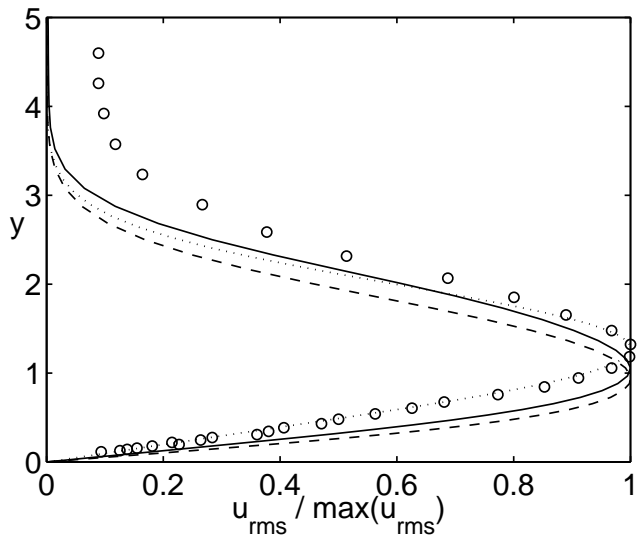}
\caption{\label{Urms} $r.m.s$ profiles for the streamwise velocity : for the edge state at $Re_{\delta^*}=500$ (continuous line) and $Re_{\delta^*}=1000$ (dashed line), the optimal streaks (dotted line), and experiments (circles). The curves have been normalized with their maximum value. }
\end{center}
\end{figure}

%--------------------------------------------------------------------------------------------------
%  comparisons with the optimal streaks and the experiments by Coupland T3A-
%--------------------------------------------------------------------------------------------------
The edge state is dominated by two streamwise vortices with an associated low-speed streak; as such, the edge-state bears a resemblance to optimal disturbances. The streamwise velocity fluctuation profile ($u^2_{rms} = \int_{xzt} (u-\bar{U})^2~ dx~dz~dt$) is presented in figure \ref{Urms}. 

The result is compared to the profile from the optimal perturbation theory (see Luchini \cite{L2000} and Andersson \emph{et al.} \cite{Andersson1}) and with experimental data by Coupland \cite{RollsRoyce} (T3A- test case). The experiment represents profiles of $r.m.s.$ velocity perturbations measured in a flat-plate boundary layer  at location $Re_{\delta^*}=1776$, perturbed by grid-generated free stream turbulence. The turbulence level is lower than 1\% at the leading edge, thus the turbulence is triggered by classical unstable Tollmien-Schlichting waves. That background turbulence is the reason why the experimental profile does not tend to zero outside the boundary layer. The wall normal mode shape of the streamwise $u_{rms}$ edge state is slightly different from the optimal perturbation or the values measured in a transitional boundary layer. In particular, the maximum value is attained closer to the wall, this can be explained by the two outer vortices, shown in the figure \ref{EdgeState_Streaks_Rolls}, which maintain the structure closer to the wall. The time averaged amplitude of $u_{rms}$ is $12\%U_\infty$ ($Re_{\delta^*}=500$) and $9\%U_\infty$ ($Re_{\delta^*}=1000$) for the edge state, which is close to the respective  values $14\%U_\infty$ and $13\%U_\infty$ observed in the turbulent case.  In addition, the $rms$ amplitude of the streamwise fluctuations is at least one order of magnitude greater than the values for the crossflow fluctuations, in agreement with the lift-up mechanism, indeed Luchini \cite{L2000} demonstrated that the amplitude ratio between the streaks and streamwise vortices scales with the Reynolds number.

\begin{figure}[!h]
\centerline{\includegraphics[width=12cm]{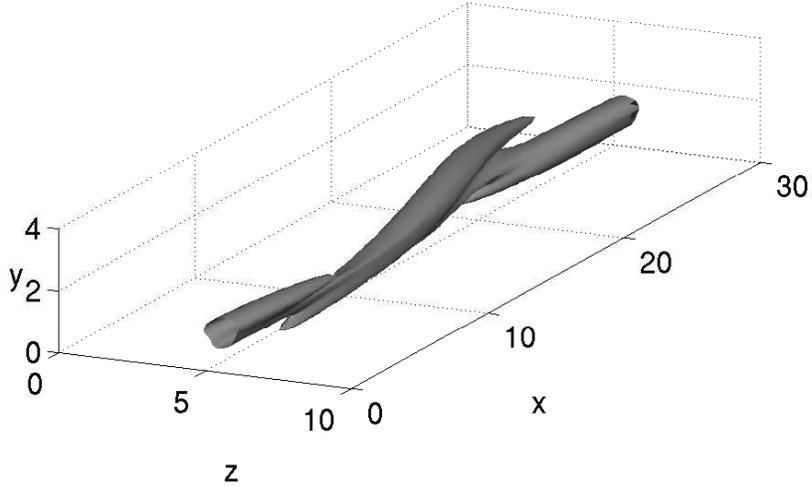}}
  \caption{Vortical structures on the separatrix at $t=2800$ for $Re_{\delta^*}=500$ and $L_x=30$.}
\label{EdgeState_Q}
\end{figure}

%--------------------------------------------------------------------------------------------------
%  hairpin or not hairpin?
%--------------------------------------------------------------------------------------------------
Figure \ref{EdgeState_Q} provides a snapshot showing of the three dimensional vortical structure identified by the Q-criterion.
The Q-criterion locates regions where rotation dominates strain in the flow, see Hunt \emph{et al.} \cite{Hunt}. Letting $S$ and $\Omega$ denote the symmetric and antisymmetric parts of the velocity gradient tensor $\nabla u$, the definition of the Q-factor is given by: 
$Q=(\Omega_{ij}\Omega_{ij}-S_{ij}S_{ij})/2$. A coherent vortex is defined as a region where $Q$ is positive.

The main structures consist of quasi-streamwise vortices, of alternating sign, in a staggered pattern located on the flanks of the low-speed streak located in the middle of the box, see also the figure \ref{EdgeState_Streaks_Rolls}. The observed sinuous structures are in agreement with the streaks of stability analysis indicating that the sinuous modes preferentially initiate the breakdown process. The overview on streaks-breakdown by Schlatter \emph{et al.} \cite{Schlatter2} also confirm the prevalence of the sinuous structures over the varicose structures for transitional flows. 

However, there exist numerous paths from laminar flow to turbulence, depending on the initial conditions, their amplitude or shape.
In particular, typical varicose structures, taking the form of hairpin or horseshoe-shaped vortices, have been observed in a transitional spatial boundary layer by Wu \& Moin \cite{Wu} and  Cherubini \emph{et al.} \cite{Cherubini}. Nonetheless, the edge state on the separatrix is found very close to the laminar state and it is characterized by fluctuations with relatively low amplitudes. Thus strong varicose structures appearing during the transition to turbulence may be considered as subsequent or secondary structures appearing on the path to turbulence, beyond the separatrix. Alternatively those structures may be on a different trajectory, initialized outside the laminar basin.

Wu \& Moin \cite{Wu} present instantaneous flow fields in both transitional and turbulent regions which are populated by hairpin vortices. The  hairpin-shaped vortices, not just quasi-symmetric or one legged hairpins, have been observed clearly in both the later parts of the by-pass transition and in the fully developed turbulent region. They also observed a peculiar bump in the Reynolds stress curves which is slowly damped downstream but still visible in the developed part of the turbulence. Wu \& Moin \cite{Wu} argued that none of the previous simulations were of a genuine spatially developing turbulent boundary layer. From a different point of view, it could be argued that the presence of hairpin vortices, in this case, is a consequence of the periodic forcing used to trigger the transition. After passage of the free-stream turbulence pocket, the streamwise flow is straightened and enters into the relatively slower-moving turbulent puff. The passage of this high-speed plug flow past the slower fluid, that resides near the wall, leads to a strong spanwise sheet of vorticity at the interface, away from the wall. The roll-up of this shear layer, through the Kelvin-Helmholtz instability, could be interpreted as the head of the hairpin vortices. This is one of the possible explanations for the production of hairpin-vortices presented by \cite{Panton}. This effect has been experimentally investigated, in a different context, by Bandyopadhyay \emph{et al.} \cite{Bandyopadhyay} for the turbulent puff in transitional pipe flow.

Cherubini \emph{et al.} \cite{Cherubini} have simulated the evolution of a nonlinear optimal impulse (localized in time and in streamwise direction) in spatially developing flow. They found the threshold amplitude above which the disturbances degenerate into a turbulent puff, suggesting the existence of relative attractors which are formed by a main hairpin vortex placed at the leading edge of the packet, followed by a population of quasi-streamwise vortices. Their simulations are realized for a super-critical Reynolds number, thus the solution evolves in between the classical path to turbulence (involving the slow increase of a T-S wave-packet) and the faster path where the T-S amplification is stimulated by the Orr mechanism (see Butler \& Farrell \cite{ButlerFarrell}). This could explain the presence of hairpin vortices which are the classical nonlinear development of the T-S instability (see Schmid \& Henningson \cite{YellowBook}).

%--------------------------------------------------------------------------------------------------
\section{Conclusion}\label{sec:conclusion}
%--------------------------------------------------------------------------------------------------
The knowledge of the edge states is important for understanding the sub-critical transition to turbulence, providing a complementary tool, beyond the linear or weakly linear approaches. At the boundary between order and chaos, we observe the features of both states. There is enough order to extract some information about the mechanisms triggering the transition, and enough nonlinearity to provide indications about the dynamics of coherent structures.  The vortical structures show a sinuous symmetry, in agreement with streak linear stability analysis. However, observations of transitional and turbulent streaks have revealed the presence of both sinuous and varicose symmetries at breakdown. One may regret the limitation of the model for the large scale structures. Because of the strong hypothesis of the model, especially the streamwise periodicity, the results cannot be extended to the large scale motions. Thus the results regarding the appearance of sinuous vortices cannot be considered as definitive and that open question deserves further investigation. Nonetheless the possibility to extend the method to unstable flows, providing that the streamwise length is short enough, opens up possibilities for the application to a wide variety of shear flows, like the mixing layer for example.

An open question concerns the possible existence of equilibrium states on the separatrix. The present challenge is to describe the flow dynamics as a trajectory in phase space meandering around typical equilibrium states (time-periodic solutions). This idea can be linked to the concept of active and inactive motion presented by Townsend \cite{Townsend}, where the exact coherent solutions provide the active part, while the residual (assumed to be uncorrelated to the coherent state) contain the inactive motion. These exact coherent solutions, with their stable and unstable manifolds, are typically low-dimensional saddle points which collectively produce a chaotic repeller that is a random switching from one unstable periodic solution to another. The complexity of these dynamics necessitates a more complex iterative method than the bisection with a single parameter governing the fluctuation amplitude. A perspective would be to pursue the method with two parameters, where the second is used to constrain the dynamics of the solution on the separatrix or to compute edge states with more than two unstable directions.

%Acknowledgements
\begin{acknowledgements}
The author wishes to thank Dr P. Jordan for the stimulating discussions on the subject.
\end{acknowledgements}

% Create the reference section using BibTeX:


\begin{thebibliography}{}

\end{thebibliography}


\begin{thebibliography}{10}

\bibitem{Spalart}
P.~R. Spalart.
\newblock Direct numerical simulation of turbulent boundary layer up to
  {R}e$_\theta$ = 1410.
\newblock {\em J. Fluid Mech.}, 187:61--98, 1988.

\bibitem{Koch}
W.~Koch.
\newblock On a degeneracy of temporal secondary instability modes in {B}lasius
  boundary-layer flow.
\newblock {\em J. Fluid Mech.}, 243:319--351, 1992.

\bibitem{ButlerFarrell}
K.M. Butler and B.F. Farrell.
\newblock Three--dimensional optimal perturbations in viscous shear flow.
\newblock {\em Phys. Fluids}, 1992:1637--1650, 4.

\bibitem{Andersson1}
P.~Andersson, M.~Berggren, and D.S. Henningson.
\newblock Optimal disturbances and bypass transition in boundary layers.
\newblock {\em Phys. Fluids}, 11:134--150, 1999.

\bibitem{L2000}
P.~Luchini.
\newblock Reynolds number independent instability of the boundary layer over a
  flat surface: optimal perturbations.
\newblock {\em J. FLuid Mech.}, 404:289--309, 2000.

\bibitem{Schmid_JFM_1992}
P.J. Schmid and D.S. Henningson.
\newblock A new mechanism for rapid transition involving a pair of oblique
  waves.
\newblock {\em Phys. Fluids}, A4:1986--1989, 1992.

\bibitem{Berlin}
S.~Berlin, A.~Lundbladh, and D.S. Henningson.
\newblock Spatial simulations of oblique transition in a boundary layer.
\newblock {\em Phys. Fluids}, 6:1949, 1994.

\bibitem{Klebanoff}
P.S. Klebanoff, K.D. Tidstrom, and L.M. Sargent.
\newblock The three-dimensional nature of boundary layer instability.
\newblock {\em J. Fluid Mech.}, 12:1--34, 1962.

\bibitem{Herbert}
T.~Herbert.
\newblock Secondary instability of boundary layers.
\newblock {\em Annu. Rev. Fluid Mech.}, 20:487--526, 1988.

\bibitem{YellowBook}
P.~J. Schmid and D.~S. Henningson.
\newblock {\em Stability and Transition in Shear Flows}.
\newblock Springer, New York, 1999.

\bibitem{Brandt2}
L.~Brandt, P.~Schlatter, and D.~S. Henningson.
\newblock Transition in boundary layers subject to free-stream turbulence.
\newblock {\em J. FLuid Mech.}, 517:167--198, 2004.

\bibitem{Andersson2}
P.~Andersson, L.~Brandt, A.~Bottaro, and D.S. Henningson.
\newblock On the breakdown of boundary layer streaks.
\newblock {\em J. Fluid Mech.}, 428:29--60, 2001.

\bibitem{Hoepffner}
J.~Hoepffner, L.~Brandt, and D.S. Henningson.
\newblock Transient growth of boundary layer streaks.
\newblock {\em J. Fluid Mech.}, 537:91--100, 2005.

\bibitem{Swearingen}
J.D. Swearingen and R.F. Blackwelder.
\newblock The growth and breakdown of streamwise vortices in the presence of a
  wall.
\newblock {\em J. Fluid Mech.}, 182:255--290, 1987.

\bibitem{Schlatter2}
P.~Schlatter, L.~Brandt, H.C. de~Lange, and D.S. Henningson.
\newblock On streak breakdown in bypass transition.
\newblock {\em Phys. Fluids}, 20:101505, 2008.

\bibitem{Schoppa}
W.~Schoppa and A.~Hussain.
\newblock Coherent structure generation in near-wall turbulence.
\newblock {\em J. Fluid Mech.}, 455:289--314, 2002.

\bibitem{Jimenez}
J.~Jim\`enez and P.~Moin.
\newblock The minimal flow unit in near-wall turbulence.
\newblock {\em J. Fluid Mech.}, 225:213--240, 1991.

\bibitem{Hamilton}
J.M. Hamilton, J.~Kim, and F.~Waleffe.
\newblock Regeneration mechanisms of near-wall turbulence structures.
\newblock {\em J. Fluid Mech.}, 287:317--348, 1995.

\bibitem{Nagata}
M.~Nagata.
\newblock Three-dimensional finite amplitude solutions in plane {C}ouette flow.
\newblock {\em J. Fluid Mech.}, 217:519--527, 1990.

\bibitem{Waleffe}
F.~Waleffe.
\newblock Three-dimensional coherent states in plane shear flows.
\newblock {\em Phys. Rev. Lett.}, 81:4140, 1998.

\bibitem{Wang}
J.~Wang, J.~Gibson, and F.~Waleffe.
\newblock Lower branch coherent states in shear flows: Transition and control.
\newblock {\em Phys. Rev. Lett.}, 98:204501, 2007.

\bibitem{Waleffe2}
F.~Waleffe.
\newblock On a self-sustaining process in shear flows.
\newblock {\em Phys. Fluids}, 9:883, 1997.

\bibitem{Cvitanovic}
P.~Cvitanovi\'{c} and J.F. Gibson.
\newblock Geometry of turbulence in wall-bounded shear flows: Periodic orbits.
\newblock {\em Phys. Scr.}, T142:014007, 2010.

\bibitem{Kawahara2}
G.~Kawahara.
\newblock Theoretical interpretation of coherent structures in near-wall
  turbulence.
\newblock {\em Fluid Dyn. Res.}, 41:064001, 2009.

\bibitem{Panton}
R.L. Panton.
\newblock Overview of the self-sustaining mechanisms of wall turbulence.
\newblock {\em Prog. in Aerospace Sciences}, 37:341--383, 2001.

\bibitem{Hof}
B.~Hof, C.W.H. van Doorne, J.~Westerweel, F.T.M. Nieuwstadt, H.~Faisst,
  B.~Eckhardt, H.~Wedin, R.R. Kerswell, and F.~Waleffe.
\newblock Experimental observation of nonlinear traveling waves in turbulent
  pipe flow .
\newblock {\em Science}, 305:1594--1598, 2004.

\bibitem{Duriez}
T.~Duriez, J.-L. Aider, and J.E. Wesfreid.
\newblock Self-sustaining process through streak generation in a flat-plate
  boundary layer.
\newblock {\em Phys. Rev. Lett.}, 103:144502, 2009.

\bibitem{Toh}
S.~Toh and T.~Itano.
\newblock A periodic-like solution in channel flow.
\newblock {\em J. Fluid Mech.}, 481:67--76, 2003.

\bibitem{Skufca}
J.~D. Skufca, J.~A. Yorke, and B.~Eckhardt.
\newblock Edge of chaos in a parallel shear flow.
\newblock {\em Phys. Rev. Lett.}, 96:174101, 2006.

\bibitem{Schneider2}
T.~M. Schneider, B.~Eckhardt, and J.~A. Yorke.
\newblock Turbulence transition and the edge of chaos in pipe flow.
\newblock {\em Phys. Rev. Lett.}, 99:034502, 2007.

\bibitem{Schneider1}
T.~M. Schneider, J.~F. Gibson, M.~Lagha, F.~De Lillo, and B.~Eckhardt.
\newblock Laminar-turbulent boundary in plane {C}ouette flow.
\newblock {\em Phys. Rev. E}, 78:037301, 2008.

\bibitem{Brandt}
L.~Brandt.
\newblock Numerical studies of the instability and breakdown of a
  boundary-layer low speed streak.
\newblock {\em Eur. J. Mech. B/Fluids}, 26:64--82, 2007.

\bibitem{Lund}
T.~S. Lund, X.~Wu, and K.~D. Squires.
\newblock Generation of turbulent inflow data for spatially-developing boundary
  layer simulation.
\newblock {\em J. Comput. Phys.}, 140:233--258, 1998.

\bibitem{Simens}
M.~P. Simens, J.~Jim\`enez, S.~Hoyas, and Y.~Mizuno.
\newblock A high-resolution code for turbulent boundary layers.
\newblock {\em J. Comp. Phys.}, 228:4218--4231, 2009.

\bibitem{SpalartYang}
P.~R. Spalart and K.~Yang.
\newblock Numerical study of ribbon induced transition in {B}lasius flow.
\newblock {\em J. Fluid Mech.}, 178:345--365, 1987.

\bibitem{Botella}
O.~Botella.
\newblock On the solution of the {N}avier-{S}tokes equations using {C}hebyshev
  projection schemes with third-order accuracy in time.
\newblock {\em Computers \& Fluids}, 26:107--116, 1997.

\bibitem{Schlatter}
P.~Schlatter and R.~\"{O}rl\"{u}.
\newblock Assessment of direct numerical simulation data of turbulent boundary
  layers.
\newblock {\em J. Fluid Mech.}, 659:116 -- 126, 2010.

\bibitem{Alfredsson}
P.~H. Alfredsson, A.~V. Johansson, J.~H. Haritonidis, and H.~Eckelmann.
\newblock The fluctuating wall‐shear stress and the velocity field in the
  viscous sublayer.
\newblock {\em Phys. Fluids}, 31:1026, 1988.

\bibitem{Orlu}
R.~\"{O}rl\"{u} and P.~Schlatter.
\newblock On the fluctuating wall-shear stress in zero pressure-gradient
  turbulent boundary layer flows.
\newblock {\em Phys. Fluids}, 23:021704, 2011.

\bibitem{Duguet}
Y.~Duguet, A.P. Willis, and R.R Kerswell.
\newblock Transition in pipe flow: the saddle structure on the boundary of
  turbulence.
\newblock {\em J. Fluid Mech.}, 613:255--274, 2008.

\bibitem{Biau}
D.~Biau and A.~Bottaro.
\newblock An optimal path to transition in a duct.
\newblock {\em Phil. Transact. Royal Soc.}, 367:529--544, 2009.

\bibitem{RollsRoyce}
J.~Coupland.
\newblock Transition modeling for turbomachinery flows, {T}3 test cases.
\newblock {\em ERCOFTAC  Bulletin}, No. 5, 1990.

\bibitem{Hunt}
J.C.R. Hunt.
\newblock Vorticity and vortex dynamics in complex turbulent flows .
\newblock {\em Transaction of CSME}, 11:21--35, 1987.

\bibitem{Wu}
X.~Wu and P.~Moin.
\newblock Direct numerical simulation of turbulence in a
  nominally-zero-pressure-gradient flat-plate boundary layer.
\newblock {\em J. Fluid Mech.}, 630:5--41, 2009.

\bibitem{Cherubini}
S.~Cherubini, P.~De Palma, J.-Ch. Robinet, and A.~Bottaro.
\newblock Edge states in a boundary layer.
\newblock {\em Phys. Fluids}, 2011:051705, 23.

\bibitem{Bandyopadhyay}
P.~R. Bandyopadhyay.
\newblock Aspects of the equilibrium puff in transitional pipe flow.
\newblock {\em J. FLuid Mech.}, 163:439--458, 1986.

\bibitem{Townsend}
A.A. Townsend.
\newblock {\em The structure of turbulent shear flows}.
\newblock Cambrige University Press, 1956.

\bibitem{footnote1}
The energy stability limit is determined with the optimal perturbation method described by Butler \& Farrell \cite{ButlerFarrell}, see also Schmid \& Henningson \cite{YellowBook}, pp. 188-193.

\bibitem{footnote2}
With the model described in the present article, a sustained turbulent flow is observed up to the minimum Reynolds number $Re_{\delta^*}=390$, which is close to the value reported by Spalart \cite{Spalart}: $Re_{\delta^*}= 400$ (\emph{i.e.} $Re_\theta=225$). Nonetheless, these values are in disagreement with Preston's estimate that the lowest Reynolds number at which fully developed turbulent flow can occur is $Re_\theta=320$, \emph{i.e.} $Re_{\delta^*}\approx 520$. Note that the latter value corresponds to the critical Reynolds number for the Tollmien-Schlichting instability. This discrepancy was discussed by Spalart \cite{Spalart}, who has pointed out a distinction between the concepts of \emph{fully developed turbulence} ($Re_{\delta^*} \ge 1000$) and the \emph{sustained turbulence} ($400 \le Re_{\delta^*} \le 1000$). In the former case, a normal log layer can be clearly distinguished while this is not obvious in the latter case; despite the fact that turbulence is observed for an arbitrary long period.

\end{thebibliography}
\end{document}